\documentclass[usenatbib]{mnras}
\usepackage{aas_macros}
\usepackage{amsmath}
\usepackage{amssymb}
\usepackage{color}
\usepackage{epsfig}
\usepackage{float}
\usepackage{graphicx}
\usepackage{latexsym}
\usepackage{morefloats}
\usepackage{natbib}
\usepackage{subfigure}
\usepackage{times}
\usepackage{auto-pst-pdf}

\newcommand{\plotone}[1]{\resizebox{0.93\hsize}{!}{\includegraphics{#1}}}
\newcommand{\plottwo}[2]{\resizebox{0.95\hsize}{!}
{\includegraphics{#1}\hspace{.5cm}\includegraphics{#2}}}

\newcommand{\Msun}{{~\rm M_\odot}}

\newcommand{\kpc}{~\rm kpc}
\newcommand{\Mpc}{~\rm Mpc}

\def\gsim { \lower .75ex \hbox{$\sim$} \llap{\raise .27ex \hbox{$>$}}}
\def\lsim { \lower .75ex \hbox{$\sim$} \llap{\raise .27ex \hbox{$<$}}}
\newcommand{\eagle}{\textsc{eagle}}

\newcommand{\simRef}{Ref-L{\small 0100}N{\small 1504}}

\newcommand{\refsec}[1]{Section \ref{#1}}

\newcommand{\reffig}[1]{Fig. \ref{#1}}

\definecolor{blue-violet}{rgb}{0.54, 0.17, 0.89}
\definecolor{purple}{rgb}{0.5, 0., 0.5}

\title[]{\parbox[b]{\linewidth}
{Evolution of LMC/M33-mass dwarf galaxies in the \textsc{eagle} simulation}}
\author[Shao et al.]
{\parbox{\textwidth}{
 Shi Shao$^{1}$\thanks{E-mail: shi.shao@durham.ac.uk}, Marius Cautun$^{1}$, Alis J. Deason$^{1}$,
 Carlos S. Frenk$^{1}$ and Tom Theuns$^{1}$ \vspace{.20cm}}\\
$^1$Institute for Computational Cosmology, Department of Physics, Durham University, South Road Durham DH1 3LE, UK \\
}

\pubyear{2018}

\begin{document}
\label{firstpage}
\pagerange{\pageref{firstpage}--\pageref{lastpage}}
\maketitle

\begin{abstract}
  We investigate the population of dwarf galaxies with stellar masses
  similar to the Large Magellanic Cloud (LMC) and M33 in the \eagle{} galaxy
  formation simulation. In the field, galaxies reside in haloes with
  stellar-to-halo mass ratios of $1.03^{+0.50}_{-0.31}\times10^{-2}$
  (68\% confidence level); systems like the LMC, which have an
  SMC-mass satellite, reside in haloes about 1.3 times more massive,
  which suggests an LMC halo mass at infall,
  $M_{200}=3.4^{+1.8}_{-1.2}\times10^{11}\Msun$ (68\% confidence
  level). The colour distribution of dwarfs is bimodal, with the red
  galaxies ($g-r>0.6$) being mostly satellites. The fraction of red
  LMC-mass dwarfs is 15\% for centrals, and for satellites this
  fraction increases rapidly with host mass: from 10\% for
  satellites of Milky Way (MW)-mass haloes to nearly 90\% for
  satellites of groups and clusters. The quenching timescale, defined
  as the time after infall when half of the satellites have acquired
  red colours, decreases with host mass from ${>}5$ Gyrs for MW-mass
  hosts to $2.5$ Gyrs for cluster mass hosts. The satellites of
  MW-mass haloes have higher star formation rates and bluer colours
  than field galaxies. This is due to enhanced star formation
  triggered by gas compression shortly after accretion. Both the LMC
  and M33 have enhanced recent star formation that could be a
  manifestation of this process. After infall into their MW-mass
  hosts, the $g-r$ colours of LMC-mass dwarfs become bluer for the
  first 2 Gyrs, after which they rapidly redden. LMC-mass dwarfs fell
  into their MW-mass hosts only relatively recently, with more than
  half having an infall time of less than 3.5 Gyrs.
\end{abstract}


\section{Introduction}
Of the multitude of galaxies in the cosmos, dwarf galaxies are the
most abundant and, at the same time, amongst the least understood.
Galaxy formation is a complex process and even more so in the case of
dwarf galaxies. For example, in the standard cosmological model, only
a small fraction of low mass haloes are occupied by galaxies. Even for
those that have a luminous counterpart, the relation between galaxy
and dark matter halo properties is an intricate one, which is shaped
by a diverse set of feedback processes \citep[see e.g. the review
of][]{Benson2010}. Here, we focus on LMC-mass galaxies (i.e. the most
massive dwarfs) and study their properties in the \eagle{}
cosmological simulation \citep{Schaye2015,Crain2015}, which resolves a
large number of such objects. These can be readily compared to
observations, where LMC-mass dwarfs can be studied out to relatively
large cosmological scales in a variety of environments
\citep[e.g.][]{Woods2007,Pozzetti2010,Tollerud2011,Geha2012,Bauer2013}. Furthermore,
the study of LMC-mass galaxies is key for understanding the formation
history of the LMC and M33, the brightest satellites of the Milky Way
(MW) and M31, respectively.

LMC-mass galaxies reside in relatively low mass haloes, of typical
mass a few times $10^{11}\Msun$ \citep{Moster2010,Guo2010}, and have a
diversity of colours and star formation rates (SFR). Large redshift
surveys, such as the Sloan Digital Sky Survey (SDSS), have revealed
that LMC-mass dwarfs have a bimodal $g-r$ colour distribution, forming
an extended blue cloud and a narrower red sequence
\citep{Strateva2001}. The fraction of red dwarfs varies with
environment: LMC-mass field galaxies are significantly bluer than
similar mass satellites \citep{Tollerud2011}. The same trend is seen
in the star formation of LMC-mass galaxies, with overdense regions
having a larger fraction of quiescent dwarfs
\citep{Wijesinghe2012}. The trend in the fraction of red and quiescent
galaxies with environment is a manifestation of quenching processes,
such as ram pressure stripping and starvation, that typically act when
galaxies reside in dense environments or become satellites of a more
massive galaxy \cite[e.g.][for a review;
\citealt{Wetzel2013,Fillingham2016,Bahe2017,Simpson2017,Fattahi2018}]{Blanton2009}.

Intriguingly, the LMC-mass satellites of the MW and M31, the LMC and
M33 respectively, have very blue colours and are actively forming
stars \citep{Munoz-mateos2007,Harris2009,Eskew2011,Tollerud2011}. For
example, the LMC is unusually blue; it lies in the ${\sim}1\%$ tail of
the SDSS $g-r$ colour distribution of galaxies of the same magnitude
\citep{Tollerud2011}, and is forming more stars than expected on
average for its stellar mass. This seems contrary to the average
expectation that satellite galaxies should have redder colours and
lower SFR, and raises the question of how efficient are MW-mass haloes
at quenching their brightest satellites. The orbital dynamics of the
LMC and M33 suggest that both these satellites were accreted recently,
typically less than 2 Gyrs ago, and are on their first orbit around
their central galaxies
\citep{Kallivayalil2006,Kallivayalil2013,Deason2015,Patel2017,Laporte2018,Cautun2018}. Furthermore,
SDSS observations find that the fraction of red satellites decreases
with host halo mass and, for MW-mass host haloes, the blue satellites
become more numerous than the red satellites
\citep[e.g.][]{Weinmann2006,Kimm2009,Guo2013,Wang2014}. Typically,
theoretical models fail to reproduce this trend, although
\citet{Sales2015} have found a good agreement between observations and
the galaxy population of the \textsc{illustris} hydrodynamic
simulation.

The LMC and M33 are peculiar in another respect: only a small fraction
of MW-mass systems are expected to host such bright
satellites. Observationally, studies of MW-like galaxies in the SDSS
\citep{Liu2011,Guo2011,Tollerud2011,Wang2012,Guo2013} and in the
Galaxy And Mass Assembly (GAMA; \citealt{Robotham2012}) surveys have
found that only about $10\%$ have satellites as bright as the
LMC. Systems that additionally have a Small Magellanic Cloud (SMC),
which observations suggest fell into the MW as a satellite of the LMC
\citep[and discussion within]{Kallivayalil2013}, are even more
rare. This result is confirmed by numerical simulations, which also
show that the probability of having an LMC satellite varies strongly
with host halo mass \citep[e.g.][]{Boylan-Kolchin2010,Busha2011}.

The importance of the LMC, and possibly M33, is also reflected in the
``satellites-of-satellites" population, which are dragged into the MW
by more massive dwarfs. For example, the SMC, and potentially a large
fraction of the dwarfs recently discovered by the Dark Energy Survey (DES) 
\citep{Jethwa2016,Sales2017}, were likely satellites of the LMC. Due
to its large total mass, with current estimates suggesting a total
halo mass of $2.5\times10^{11}\Msun$
\citep{Penarrubia2016,Cautun2018}, the LMC is expected to have
contributed up to $30\%$ of the current MW satellite population
\citep{Deason2015,Shao2017}.

In this paper, we study the properties of a large sample of LMC-mass
galaxies in the \eagle{} galaxy formation
simulation. 
\eagle{} is ideal for this study since it reproduces a range of key
observables, such as the galaxy stellar mass function, cosmic star
formation history, and galaxy sizes, metallicities, gas fractions and
morphologies across a wide range of masses and different redshifts
\citep{Furlong2015,Lagos2015,Schaye2015,Trayford2015}. LMC-mass dwarfs
are resolved in \eagle{} with about 1000 or more star particles, which
allows for a robust characterization of their morphology , SFR and
colour distribution. We probe how these properties vary according to
environment, i.e. field versus satellite galaxies, and, in particular,
we focus on LMC-mass satellites in MW-mass host haloes, with the goal
of interpreting the observed properties and evolution of the LMC and
M33.

The paper is organized as follows. Section~\ref{sect:simul} reviews the simulations used in this work and describes our sample selection; Section~\ref{sect:result} presents our results on the statistics of field and satellite LMC-mass dwarfs; Section~\ref{sect:MW} discusses the implications of our findings in the context of LMC-like satellites of MW-mass haloes; we conclude with a short summary and discussion in Section~\ref{sect:conclusions}.

\section{Simulations and methods}
\label{sect:simul}
We make use of the main cosmological hydrodynamical simulation
(labelled \simRef{}) performed as part of the \eagle{} project
\citep{Schaye2015, Crain2015}. Using a periodic cube of $100\Mpc{}$
side length, \eagle{} follows the evolution of $1504^3$ dark matter
particles, and an initially equal number of gas particles. The dark
matter particle mass is $9.7\times 10^6 \Msun$, and the initial gas
particle mass is $1.8\times 10^6 \Msun$. \eagle{} uses a \textit{Planck}
cosmology \citep{Planck2014} with cosmological parameters:
$\Omega_{\rm m}=0.307, \Omega_{\rm b}=0.04825,\Omega_\Lambda=0.693,
h=0.6777,\sigma_8=0.8288$ and $n_{\rm s}=0.9611$. 

The simulation was performed using a modified version of the
\textsc{gadget} code \citep{Springel2005}, which includes
state-of-the-art smoothed particle hydrodynamics methods
\citep{DallaVecchia2012,Hopkins2013,Schaller2015a}. The baryonic
physics implementation accounts for a multitude of processes relevant
to galaxy formation, such as element-by-element cooling using the
\citet{Wiersma2009a} prescription, stochastic star formation with a
metallicity dependent threshold \citep{Schaye2004}, thermal energy
feedback associated with star formation \citep{DallaVecchia2012}, and
the injection of hydrogen, helium and metals into the interstellar
medium from supernovae and stellar mass loss
\citep{Wiersma2009b}. Each star particle corresponds to a single
stellar population with a \citet{Chabrier2003} initial mass
function. Supermassive black holes grow through mergers and accretion
of low angular momentum material
\citep{Springel2005a,Rosas-Guevara2015,Schaye2015} and, in the
process, inject thermal energy into the surrounding gas
\citep{Booth2009,DallaVecchia2012}. The \eagle{} subgrid models were
calibrated to reproduce three present day observables
\citep{Crain2015, Schaye2015}: the stellar mass function, the
distribution of galaxy sizes, and the relation between supermassive
black hole mass and host galaxy mass. For a more detailed description,
we refer the reader to \citet{Schaye2015}.

We make use of the $z=0$ \eagle{} halo and galaxy catalogue
\citep{McAlpine2016}. Haloes are initially 
identified using the
friends-of-friends (FOF; \citealt{Davis1985}) algorithm with a linking
length $0.2$ times the mean interparticle separation. The resulting FOF
groups were further split into gravitationally bound substructures
using the \textsc{subfind} code \citep{Springel2001,Dolag2009}, which
was applied to the full matter distribution (dark matter, gas and
stars) associated with each FOF group. The main halo is determined by
the subhalo that contains the most bound particle, while the remaining
subhaloes are classified as satellites. The stellar distribution
associated with the main subhalo is identified as the central
galaxy. The main haloes are characterized by the mass, $M_{200}$, and
radius, $R_{200}$, that define an enclosed spherical overdensity of
$200$ times the critical density. The position of each galaxy, for
both centrals and satellites, is given by their most bound particle. 

\reffig{fig:Star_m200} presents the relation in \eagle{} between the
stellar masses of central galaxies and the mass of their host
haloes. We do not show satellites since their subhalo mass, which
\textsc{subfind} defines as the bound mass within the tidal radius,
varies depending on the position of the object along its orbit. The
figure shows that the stellar and halo masses are correlated, albeit
with a large scatter. The scatter, while small at large masses,
increases significantly for low mass haloes.

\subsection{Sample selection}
\begin{figure}
	\plotone{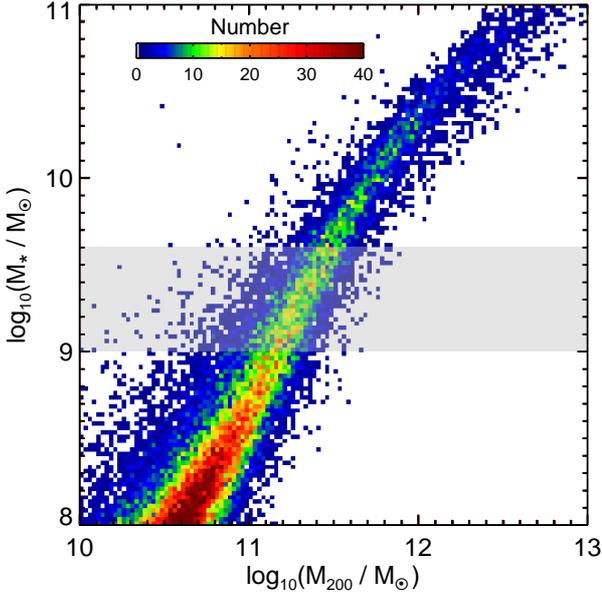}
	\caption{ The relation between stellar mass, $M_{\star}$, and
          total halo mass, $M_{200}$, for central galaxies in the
          \eagle{} simulation. The colours indicate the number of
          galaxies in each halo and stellar mass bin (see legend). The
          grey shaded region shows galaxies with stellar masses in the
          range $1-4\times 10^{9} \Msun$, which corresponds to our
          sample of field LMC-mass dwarfs (we also select LMC-mass
          satellites, which are not shown in this diagram).}
	\label{fig:Star_m200}
    \vspace{-.3cm}
\end{figure}
We select LMC-mass dwarfs by requiring that they have a stellar mass
in the range, $M_\star \in [1, 4] \times 10^{9}\Msun$, which is
motivated by the following. First, due to uncertainties in the stellar
mass to light ratio, the LMC stellar mass is somewhat uncertain, with
mass estimates spanning the range $1.5-2.7\times 10^{9}\Msun$
\citep[e.g.][]{van_der_Marel2002,McConnachie2012}. Secondly, to have
good statistics we need a large sample of LMC-mass dwarfs, and thus a
mass range as wide as possible. A typical LMC-mass dwarf in \eagle{}
is resolved with a thousand or more star particles and with hundreds
of gas particles, which allows for a robust quantification of its
present day properties as well as its formation history.

We split our LMC-mass dwarfs into two categories: (1) the satellite
galaxy sample, which consists of LMC-mass dwarfs within a radius,
$R_{\rm 50}$, from a more massive halo, and (2) the field galaxy
sample, which comprises central galaxies that are not within 
distance $R_{\rm 50}$ from a more massive halo. The $R_{\rm 50}$
radius defines an enclosed spherical overdensity of $50$ times the
critical density (it is approximatively $2^{2/3} \times R_{\rm 200}$).
We choose this bounding radius because MW studies typically
take $300\kpc$ as the Galactic halo radius, which for a MW halo mass
of $10^{12}\Msun$, corresponds to $R_{50}$. In \eagle{}, we find 3774
field\footnote{We use the term ``field" to refer to galaxies that are
not satellites of other galaxies.} galaxies and 2551 satellite
galaxies. The sample of field LMC-mass dwarfs is highlighted in
\reffig{fig:Star_m200} using a grey shaded region, which corresponds
to the stellar mass selection criteria.

We further select a subset of LMC-mass satellites that reside in MW-mass haloes, which we define as any host halo with a mass in the range, $M_{\rm 200} \in [0.5, 2] \times 10^{12}\Msun$ \citep{Cautun2014a,Wang2015,Penarrubia2016}. We find 381 LMC-mass dwarfs within $R_{50}$ of our MW-mass halo sample, with the MW-mass hosts having a median halo mass $\approx 1.0 \times 10^{12}\Msun$ and a median $R_{50} \approx 313\kpc.$

To study the evolution of LMC-mass satellites, we make use of the
\eagle{} snipshots, which are finely spaced (about every 70 Myrs)
simulation outputs that allow us to trace the orbits of satellites
with very good time resolution. We define the infall time for each
dwarf as the time when it first crosses $R_{\rm 50}$ of the progenitor
of its $z=0$ host halo.

\vspace{-.3cm}
\subsection{Galaxy morphology and colour}
\label{sect:methods_cm}
To quantify the morphology of LMC-mass dwarfs, we divide the stellar
mass of galaxies into two components: spheroid and disc, which we
identify using the procedure of \citet[][see also
\citealt{Scannapieco2009,Crain2010,Sales2012}]{Abadi2003}. We
calculate the circularity parameter of each star,
$\epsilon=j_z/j_{\rm circ}(E)$, defined as the ratio between the
component of the specific angular momentum perpendicular to the disc,
$j_z$, and that for a circular orbit with the same total energy,
$j_{\rm circ}(E)$. The disc direction is given by the angular momentum
of all the star particles within twice the half stellar mass radius,
$r_{h}$. If we assume that the spheroidal component of each galaxy is
fully velocity dispersion dominated, then the bulge mass corresponds
to twice the mass of the stars with $\epsilon<0$. Note that
$\epsilon<0$ corresponds to counter-rotating stars, i.e. stars for
which the scalar product between the stars' angular momentum and that
of the disc is negative.

We take the galaxy colours calculated by \citet{Trayford2015}, which
are based on the \textsc{galaxev} population synthesis models of
\citet{Bruzual2003}. The colours are estimated by modelling the
stellar populations of \eagle{} star particles, which represent a
simple stellar population with a \citet{Chabrier2003} initial mass
function, taking into account their ages and metallicities. The galaxy
spectra were summed over all the stellar particles within a spherical
aperture of $30\kpc$ and convolved with the colour filter response
function. Here, we take the colour of each galaxy from the intrinsic
$g-r$ colour without dust extinction. \citeauthor{Trayford2015} showed
that these colours are in broad agreement with observational data and
that, in particular, \eagle{} produces a red sequence of passive
galaxies and a blue cloud of star-forming galaxies.

\vspace{-.3cm}
\section{General properties of LMC-mass dwarfs}
\label{sect:result}
We now study general properties, such as halo mass, morphology,
colour and star formation rate (SFR) of LMC-mass dwarf galaxies. In
particular, we focus on differences between the populations of field
dwarfs and satellite galaxies, with emphasis on satellites around
MW-mass host haloes.

\vspace{-.3cm}
\subsection{Halo mass}
\label{sec:result:halo_mass}
We start by characterizing the \eagle{} haloes that host the LMC-mass
dwarfs. From \reffig{fig:Star_m200}, we find that the typical field
LMC-mass galaxy resides in a halo with a total mass of
${\sim}2\times10^{11}\Msun$, but the relation is characterized by
significant scatter. Most striking are the handful of objects with the
same stellar mass as the LMC which reside in a few
$\times10^{10}\Msun$ mass haloes. These are not satellites, since
\reffig{fig:Star_m200} shows only central LMC-mass dwarfs, and are
likely ``backsplash'' galaxies which were, at least for some period of
time, satellites around more massive host haloes and thus were tidally
stripped \citep{Moore2004}. We have checked that all the LMC-mass
dwarfs residing in haloes with $M_{200}<10^{10.5}\Msun$ are backsplash
galaxies. The scatter in the stellar-to-halo mass relation for
LMC-mass dwarfs is larger than for more massive galaxies, but is
significantly smaller than for lower mass dwarfs
\citep{Sawala2015}. For LMC-mass galaxies, a large fraction of the
scatter is due to haloes having different concentrations and binding
energies \citep{Matthee2017}. Higher concentration objects, which
typically formed earlier, have more time to form stars and experience
less efficient feedback since they are more tightly bound.

\begin{figure}
	\plotone{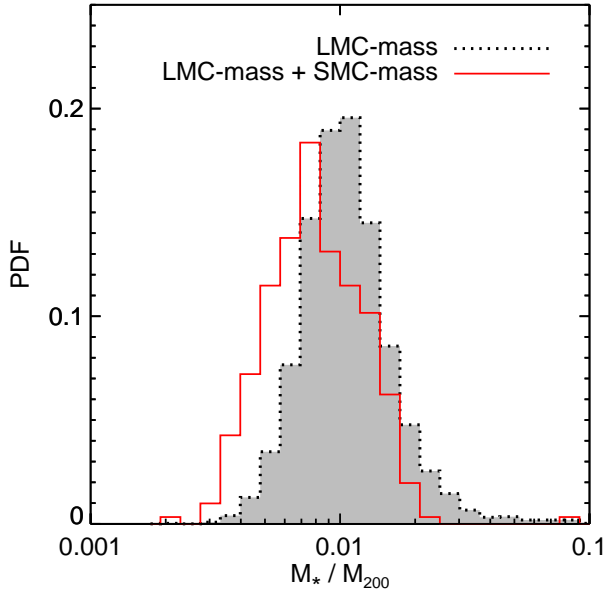}
	\caption{ The distribution of stellar-to-halo mass ratios at
          $z=0$ in the \eagle{} simulation for \textit{field} LMC-mass
          dwarfs. We show the distribution for all LMC-mass galaxies
          (dotted line) and for those that have an SMC-mass satellite
          (solid line). Having an SMC-mass satellite biases the
          LMC-mass dwarfs towards 1.3 times higher halo masses. 
	}
    \label{fig:Pdf_mf}
    \vspace{-.3cm}
\end{figure}
We further study the stellar-to-halo mass ratio, $M_\star/M_{200}$, in
\reffig{fig:Pdf_mf}, where we present the probability distribution
function (PDF) of $M_{\star}/M_{\rm 200}$ for LMC-mass field
dwarfs. The distribution is peaked at a value of
$1.03^{+0.50}_{-0.31}\times 10^{-2}$ (68\% confidence limit), with a
sharp drop-off on both sides; this is in agreement with results from
SDSS abundance matching models, although the dispersion of the
distribution is larger than the $0.15-0.20$ dex scatter typically
assumed in these models \citep{Moster2010,Guo2010}. 

Both the LMC and M33 are predicted to have been accreted recently
\citep{Patel2017}, 1.5 and 0.4 Gyrs ago, respectively. We can assume
that their halo masses at infall are likely similar to their present
day masses, under the assumption that they are not significantly
tidally stripped. Also, both galaxies are unlikely to have increased
their stellar masses by more than $10\%$ since infall,
so their infall stellar masses roughly correspond
to their present day masses. Finally, for the same LMC-mass galaxy
selection criteria, \eagle{} predicts the same $M_{\star}/M_{\rm 200}$
ratio for centrals at a slightly higher redshift,
e.g. $z=0.2$. Thus, we can use the $M_{\star}/M_{\rm 200}$
distribution in \eagle{} to estimate the LMC halo mass at infall. Taking an LMC
stellar mass of $2.7\times10^{9}\Msun$ \citep{van_der_Marel2002}, we
estimate the LMC total mass to be
$2.6^{+1.1}_{-0.9} \times 10^{11}\Msun$ (68\% confidence limit),
which is in agreement with the dynamical mass estimates of
$2.5\pm 0.8\times10^{11}\Msun$ by \citet{Penarrubia2016}. M33 has a
similar stellar mass, $3.0\times10^9\Msun$ \citep{McConnachie2012},
and thus is expected to reside in a similar mass halo.

Observations indicate that the two brightest MW satellites, the LMC
and SMC, were accreted as part of the same group. Does this
observation bias the LMC mass estimates? To answer this question, we
proceed by identifying field LMC-mass galaxies that have an SMC-mass
satellite. The SMC has a stellar mass roughly one third of the LMC
\citep{McConnachie2012}, so we define SMC-mass satellites as any
objects with a stellar mass $\sim0.2$ times or higher than that of its
central LMC-mass galaxy. The binary LMC-SMC analogues reside in
significantly more massive haloes for their stellar mass (see dashed
curve in \reffig{fig:Pdf_mf}), with this sample having
$M_{\star}/M_{\rm 200}=0.79^{+0.45}_{-0.27}\times 10^{-2}$ (68\%
confidence limit). Thus, the LMC halo is a factor of 1.3 times more
massive than that of the typical LMC-mass dwarf and likely contributes
a significant fraction of the mass of the MW halo ($\sim 20-40\%$ for
a MW halo mass of $1\times10^{12}\Msun$).

\vspace{-.3cm}
\subsection{Morphology}
\begin{figure}
	\plotone{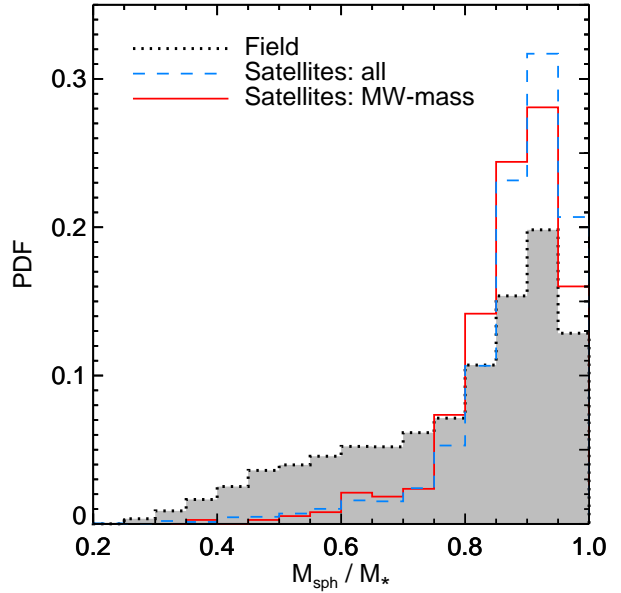}
	\caption{ The morphology distribution of LMC-mass dwarfs. The
          morphology, defined as the ratio of the spheroid-to-total
          stellar mass within twice the half mass radius, is
          calculated by a dynamical decomposition into two components:
          disc and spheroid. The three lines show the distribution for
          LMC-mass dwarfs in the field (dotted line) and for
          satellites around all hosts (dashed line) and around MW-mass
          hosts (solid line). 
	}
	\label{fig:Pdf_epsilon}
    \vspace{-.4cm}
\end{figure}
\reffig{fig:Pdf_epsilon} shows the distribution of the
spheroid-to-total stellar mass ratio, $M_{\rm sph}/M_{\star}$, from a
kinematic decomposition of each LMC-mass galaxy into bulge and disc
components, as described in \refsec{sect:methods_cm}. Most of the
field LMC-mass objects are bulge dominated; over $60\%$ of the sample
have $M_{\rm sph}/M_{\star} > 0.8$, which indicates that these
galaxies are typically spherical and are velocity dispersion
supported. A significant fraction ($\sim 20\%$) of field dwarfs have
$M_{\rm sph}/M_{\star} < 0.6$, which indicates that they have
significant ordered rotation. In contrast, there are very few LMC-mass
satellites (3\%) that show a disc-like morphology; most objects have
$M_{\rm sph}/M_{\star} \approx 0.9$, and are thus largely dominated by
their bulge. This is in qualitative agreement with visually classified
morphologies in observations, where the fraction of early type
galaxies increases in denser environments
\citep{Dressler1980,Postman1984,Bamford2009}.

The morphology distribution of LMC-mass satellites around MW-mass
haloes is very similar to that of satellites around all hosts. As we
show later in \reffig{fig:Pdf_infall}, roughly half of the LMC-mass
satellites of MW-mass haloes were accreted recently ($<3.5$ Gyrs), and
thus the lack of disc-like morphologies in LMC-mass satellites,
compared to their field counterparts, is puzzling. It suggests that in
\eagle{}, once some of the galaxies become a satellites, they undergo
a rapid morphological transformation. The lack of disky satellites is
also puzzling when comparing to observations, which find a larger
fraction of late-type galaxies \citep[e.g.][]{Bamford2009}. A similar
discrepancy is present when comparing with the two Local Group
satellites, the LMC and M33, which have disc-like morphologies. M33 is
visually classified as a disc and the LMC, while visually classified
as an irregular galaxy, is kinematically dominated by ordered rotation
more akin to that of a disc galaxy \citep{van_der_Marel2002}. 
The differences in the morphologies of dwarf galaxies between 
\eagle{} and observations are unlikely to be due to resolution effects: 
LMC-mass dwarfs in \eagle{} are resolved with ${\sim}1000$ particles. 
To verify this we compared the morphology of LMC-mass dwarfs in two simulations from 
the \eagle{} project (with side-length 25~Mpc), one at the fiducial resolution 
and the other at 8 times better mass resolution; the distribution of spheroid-to-total
stellar mass ratios are approximately the same in the two simulations. Furthermore, 
\citet{Benitez-Llambay2018} demonstrated that the galaxy formation model 
employed in \eagle{} is able to reproduce adequately the structure of 
disc-like galaxies. We note that a direct
comparison of our results to observations is difficult because the 
disc/spheroid kinematic decomposition we use in the simulations
differs from the customary photometry-based methods used in observational
studies. Indeed, the correspondence between these two methods has
significant scatter, and photometric decomposition methods tend to
estimate lower bulge-to-total ratios, especially for low mass galaxies
\citep{Abadi2003,Okamoto2005,Scannapieco2010,Bottrell2017}. 

\vspace{-.3cm}
\subsection{Star formation rate}
\begin{figure}
	\plotone{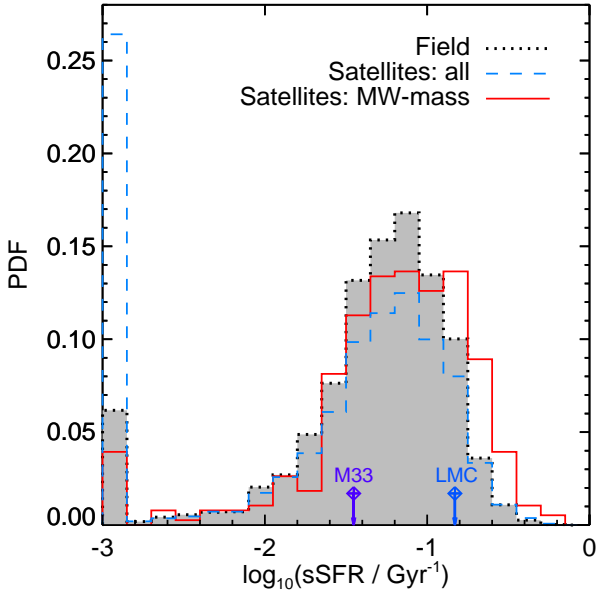}
	\caption{ The specific star formation rate (sSFR) distribution
          of LMC-mass dwarfs. The lines correspond to dwarfs
          identified in the field (dotted), satellites around all
          hosts (dashed) and satellites around MW-mass haloes
          (solid). The galaxies that have zero or extremely small sSFR
          are all grouped together in the left-most bin. The two
          vertical arrows indicate the observed values for the LMC and
          M33 satellite galaxies. 
    }
	\label{fig:Pdf_sfr}
    \vspace{-.3cm}
\end{figure}
The distribution of specific SFR (sSFR), $\dot{M}_\star/M_\star$, of
\eagle{} LMC-mass galaxies is given in \reffig{fig:Pdf_sfr}. The
figure shows the well known sSFR bimodality
\citep[e.g.][]{Wijesinghe2012}, with a mode that consists of
star-forming galaxies with sSFR$\sim0.06\Msun\rm{yr}^{-1}$ and a
second subsample of quiescent galaxies with none, or very little,
ongoing star formation. Quantitatively, the sSFR of star-forming
galaxies is a factor of two below observed values \citep[e.g. the GAMA
sample of][]{Bauer2013}, which is due to the overall SFRs at $z=0$ in
\eagle{} being too low \cite[for a more detailed analysis
see][]{Furlong2015}. However, this does not affect our conclusions
since our goal is to compare the differences between various dwarf
galaxy samples. The fraction of quiescent dwarfs becomes most
pronounced for the sample of satellites around all hosts and is a
manifestation of the star formation quenching processes acting on
satellite galaxies. For the star-forming population, the distribution
of sSFR for the field and all-satellite samples is roughly the
same, except for the normalization, in agreement with observational
studies \citep{Wijesinghe2012}. It suggests that once quenching
starts, it is a rapid process with a short time interval between the
stage of forming stars like a field galaxy and becoming fully
quiescent. This fits with the expectation that ``strangulation", 
the process of halting the supply of cold gas, is the main
quenching process \citep{Cole2000,Peng2015}.

It is worth noting that satellite galaxies are not always quenched,
and, at least for some time, their star formation can even be
\textit{enhanced}. This is clearly seen in the sample of LMC-mass
satellites around MW-mass haloes, which has a smaller fraction of
quiescent objects and for which the sSFR distribution of the
star-forming sample is shifted towards higher values. Indeed, enhanced
star formation may currently be taking place in the LMC, whose current
SFR is twice its mean value over the last 2 Gyrs \citep{Harris2009}.

\vspace{-.3cm}
\subsection{Colours}
\begin{figure}
 	\plotone{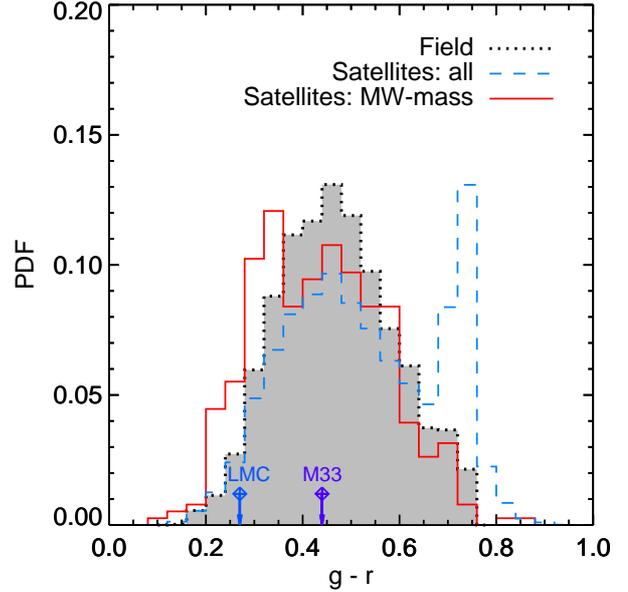}
 	\caption{ The distribution of $g-r$ colour for LMC-mass dwarfs found in the field (dotted), satellites around all hosts (dashed) and satellites around MW-mass host haloes (solid).
 	}
 	\label{fig:Pdf_color}
    \vspace{-.3cm}
\end{figure}
\reffig{fig:Pdf_color} presents the distribution of $g-r$ colours for 
our sample of LMC-mass dwarfs. While the field galaxies are well
characterized by a unimodal distribution, the all-satellite sample
is bimodal, with a subgroup of blue dwarfs peaking at
$g-r=0.45$, and a subgroup of red dwarfs peaking at $g-r=0.75$. The
\eagle{} distribution of intrinsic colours is a good match to observations
\citep[e.g. see][]{Taylor2015}, and is an even better match when 
using a dust obscuration model that depends on gas fraction
and metallicity \citep[here we use the no-dust model; for details
see][]{Trayford2015}. The LMC-mass satellites around MW-mass haloes
have bluer colours than both the field and the all-satellites samples,
and, furthermore, do not show a second ``red'' peak. As we will
discuss shortly, the subgroup of red dwarfs mainly consists of
satellites of rich groups and clusters, with $M_{\rm 200} >
1\times10^{13}$. 

The LMC has a colour $(g-r)_{\rm LMC}=0.27$ \citep{Eskew2011}, which
puts it in the tail of the field and all-satellites colour
distribution \citep[for a comparison with the $g-r$ distribution in
observations, see][]{Tollerud2011}. However, when compared to the
colour distribution of satellites around MW-mass hosts, the LMC is
no longer an outlier (10\% of \eagle{} satellites are bluer than the
LMC). M33 has a slightly redder colour, with $(g-r)_{\rm M31}=0.44$
\citep{Tollerud2011}, which is typical of a field galaxy that has been
recently accreted onto the M31 halo \citep{Patel2017}. 

\vspace{-.2cm}
\subsection{Dependence on host halo mass}
\begin{figure}
 	\plotone{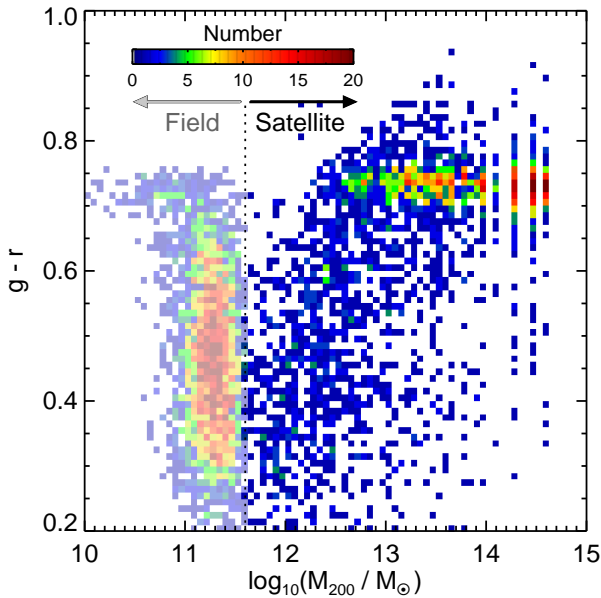}
 	\caption{ The distribution of $g-r$ colour for LMC-mass dwarfs
          as a function of halo mass. For field galaxies, the halo
          mass is that of their own halo and corresponds to the region
          left of the vertical dotted line. For satellite galaxies,
          the halo mass corresponds to that of their host haloes. The
          dotted vertical line at $M_{200}=10^{11.6}\Msun$
          approximately separates the field from the satellite population.
 	}
 	\label{fig:Color_M200}
    \vspace{-.3cm}
\end{figure}
\begin{figure}
 	\plotone{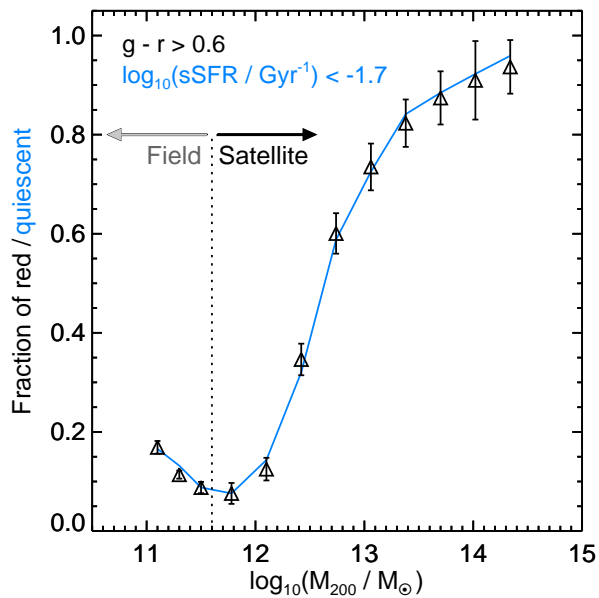}
 	\caption{ The fraction of red / quiescent LMC-mass dwarfs as a
          function of halo mass. The vertical dotted line separates
          the field galaxies (to the left) from the satellites (to the
          right). For field galaxies, the halo mass is that of their
          own halo, while for satellites, the halo mass is that of
          their host haloes. The symbols show the red fraction defined
          as galaxies with $g-r>0.6$. The line shows the quiescent
          fraction defined as dwarfs with
          $\log(\rm{sSFR}/\rm{Gyr^{-1}}) < -1.7$. The error bars
          represent the $1\sigma$ bootstrap uncertainties and are
          roughly the same for both fractions.
 	}
 	\label{fig:Pdf_red}
    \vspace{-.3cm}
\end{figure}
In order to understand the processes that shape the colour
distribution of LMC-mass galaxies, we now explore the dependence on
host halo mass. \reffig{fig:Color_M200} shows the $g-r$ colour as a
function of the halo mass for centrals and of the host halo mass for
satellites. Most field LMC-mass dwarfs have $M_{200}<10^{11.6}\Msun$,
and this mass threshold is shown as the dotted vertical line in the
figure. The LMC-mass centrals can be broadly divided into two
categories. First, there are the objects with
$M_{200}<10^{10.6}\Msun$. These have a very low halo mass for their
stellar mass and mainly correspond to backsplash galaxies (see
discussion in \refsec{sec:result:halo_mass}). Tracing the merger tree
of these objects reveals that all of them were, at some time in the
past, part of a massive, $M_{200}>10^{13}\Msun$, host halo. Secondly,
there is the main population of LMC-mass galaxies, characterized by
halo masses, $10^{10.7}\Msun<M_{200}<10^{11.6}\Msun$. While these
galaxies are predominantly blue, with a broad peak at $g-r=0.45$, the
distribution has a red tail, with a 15\% fraction of red, $g-r>0.6$,
central galaxies (see \reffig{fig:Pdf_red}). This population of
passive central galaxies could be the result of self-quenching or
mostly consist of backslash galaxies. To identify the main process,
we followed the merger tree of all red LMC-mass centrals to identify
the fraction that were satellites at any point during their formation
history. We find that at most $35\%$ of them were satellites in the
past, suggesting that the dominant process for producing LMC-mass red
centrals is self-quenching.

The colour distribution of LMC-mass satellites shows a distinct trend
with the mass of their host halo. Due to the limited volume of
\eagle{}, there are only a few haloes more massive than $10^{14}\Msun$
and each vertical strip at those masses corresponds to the satellites
in each of those hosts. Note that the average number of LMC-mass
satellites per host varies strongly with their host halo mass. A
cluster with $M_{\rm 200}\sim10^{14}$ has on average around 30
LMC-mass dwarfs, whereas only 1 out of 5 haloes with
$M_{\rm 200}\sim10^{12}$ has an LMC-mass dwarf. With a few exceptions,
there are hardly any red satellites in haloes with
$M_{200}<10^{12}\Msun$, and most satellites of ${\sim}10^{12}\Msun$
haloes are blue (defined as $g-r<0.6$). The fraction of red satellites
rapidly increases with higher host mass, with red LMC-mass dwarfs
becoming dominant in hosts more massive than $10^{12.6}\Msun$ (see
also \reffig{fig:Pdf_red}).

To quantify how many LMC-mass dwarfs are passive, we split the
population into red and and blue galaxies, according to whether 
$g-r>0.6$ or $g-r<0.6$, respectively. The fraction of red galaxies as
a function of host halo mass is shown in \reffig{fig:Pdf_red} and,
as \reffig{fig:Color_M200}, it combines in one plot both
field and satellite galaxies. We find that few field galaxies are red
(${\sim}15\%$ on average) and that the field red fraction shows a
small, but statistically significant, trend with halo mass: an LMC-mass
dwarf is slightly more likely to be red if it resides in a lower mass
halo. This trend is driven by backsplash galaxies, which, on average, 
are both redder and, due to tidal stripping, have lower halo 
masses. Interestingly, the fraction of red galaxies does not show any
discontinuity as halo mass increases and we switch from centrals to
satellites in low mass hosts. Furthermore, this transition region is
where we find the smallest fraction of red dwarfs. As the host halo
mass increases, we find a larger fraction of red satellites, with most
(${\sim}90\%$) of LMC-mass galaxies in clusters
($M_{200}>10^{14}\Msun$) having red colours. 

\reffig{fig:Pdf_red} also shows the fraction of quiescent galaxies,
which are defined as those with specific star formation rates,
${\rm sSFR} < 0.02/{\rm Gyr}$. The quiescent fraction is roughly equal
to the red fraction, and both show the exact same dependence on
mass. While most quiescent dwarfs have red colours, this is not the
case for every galaxy, with some having low sSFR and blue colours and
vice versa. This is due to the sSFR being a measure of instantaneous
star formation, while the colour is sensitive to the integrated recent
star-formation history.

The results presented in \reffig{fig:Pdf_red} are consistent with
observations, which report that it is extremely rare to find field
dwarfs with no active star formation \citep[e.g.][]{Geha2012}. The
observations also support the trend with host halo mass: most LMC-mass
satellites around faint centrals are blue, whereas most satellites in
rich groups and clusters are red
\citep[e.g.][]{Weinmann2006,Wang2014,Sales2015,Geha2017,Wang2018}. 

\begin{figure}
    \vspace{-0.2cm}
 	\plotone{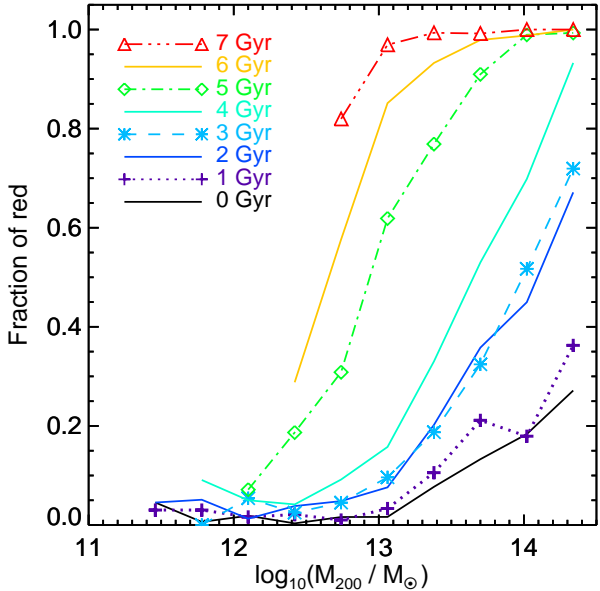}
 	\caption{ The fraction of red LMC-mass satellites as a
          function of their host halo mass. The curves show the red
          fraction at various times after infall into the host halo,
          with $t=0$ corresponding to the time of infall. The fraction
          is shown only for $M_{200}$ bins with 10 or more LMC-mass
          dwarfs. Thus, lines corresponding to infall times of 6 Gyrs
          or more do not extend down to low $M_{200}$ values
          ($\lesssim 10^{12.5}M_\odot$). 
 	}
 	\label{fig:Pdf_red_evo}
    \vspace{-0.3cm}
\end{figure}
\begin{figure}
    \vspace{-0.2cm}
 	\plotone{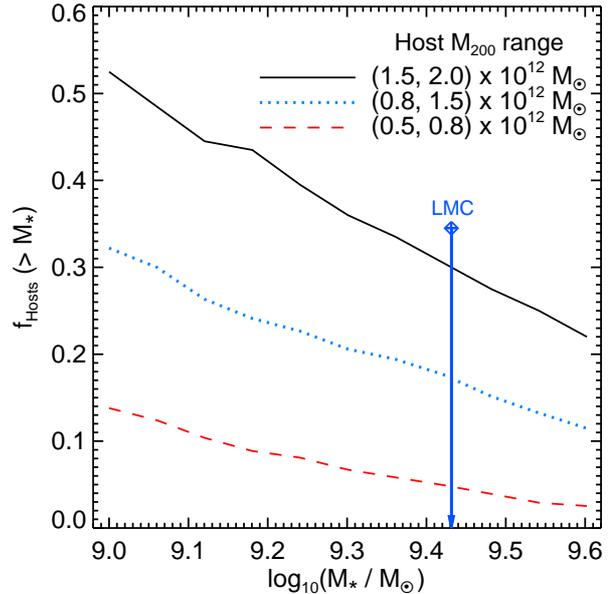}
 	\caption{ The fraction of hosts that have at least one LMC-mass satellite dwarf as a function of the stellar mass, $M_{\star}$, of the satellite. The various lines correspond to different host halo masses: $0.5-0.8$ (dashed), $0.8-1.5$ (dotted) and $1.5-2.0$ (solid) $\times 10^{12}\Msun$. Satellites are defined as galaxies within a distance, $R_{\rm 50}$, from a more massive halo. The vertical arrow indicates the \citet{van_der_Marel2002} LMC stellar mass estimate of $2.7\times 10^{9}\Msun$.
 	}
 	\label{fig:Pdf_mw}
    \vspace{-.3cm}
\end{figure}
\reffig{fig:Pdf_red_evo} presents the fraction of red satellites at
various times after infall as a function of their host halo mass. At
infall, which corresponds to $t=0$ Gyrs in the figure, most LMC-mass
satellites are blue; the only exception are the high mass haloes,
$M_{200}>10^{13}\Msun$, which accrete a non-negligible fraction of red
dwarfs. Most of these dwarfs correspond to preprocessed satellites,
which, before falling into their $z=0$ host, were satellites of
another halo \citep{McGee2009,Wetzel2013,Hou2014}. Higher mass haloes
accrete, on average, more haloes with $M_{200}\gtrsim 10^{12}\Msun$,
which can host LMC-mass satellites themselves, and thus accrete more
preprocessed LMC-mass dwarfs. 

The results shown in \reffig{fig:Pdf_red_evo} can be used to estimate
the quenching timescale for LMC-mass dwarfs as a function of their host
halo mass. For this, we follow, at fixed halo mass, the change in the
red fraction as a function of time after infall. Hosts with masses,
$M_{200}\sim 10^{12}\Msun$, have a very slowly increasing red fraction
such that, even 5 Gyrs after infall, only ${\sim}5\%$ of LMC-mass
satellites are red. Thus, these hosts have very long timescales for
quenching LMC-mass dwarfs. This is in good agreement with SDSS
observations that predict quenching timescales larger than 9 Gyrs
\citep{Wheeler2014}, and with the trends observed in the Local Group,
where the quenching time increases rapidly with the satellite stellar
mass \citep{Wetzel2015c,Fillingham2015,Simpson2017}. These long quenching times
suggest that starvation is the main quenching process, with satellites
not being able to accrete new gas. For example, the LMC had an average
SFR of $0.2\Msun~\rm{yr^{-1}}$ over the past 2 Gyrs and, given that it
has an HI gas mass of at least $0.5\times 10^9\Msun$, it can keep
forming stars at the same rate for at least another 2.5 Gyrs. By then,
the LMC would have orbited the MW for about 4 Gyrs, which is around
the time when it will merge with the MW \citep{Cautun2018}.

For hosts more massive than the MW, the quenching timescales decrease
rapidly. For example, half of the LMC-mass satellites of hosts with
masses, $M_{200}\sim 10^{13}\Msun$, are already red 5 Gyrs after
infall. For cluster mass haloes, $M_{200}\sim 10^{14}\Msun$, the
quenching is even more rapid, with half of their dwarfs being red 
$2-3$ Gyrs after infall. This is in agreement with the SDSS based
quenching timescales derived by \cite{Wetzel2013}, who also found that
quenching progresses faster in more massive haloes. This indicates
that the dominant quenching process varies with host halo mass, from
starvation in the case of MW-mass hosts to ram pressure stripping for
cluster mass hosts. The latter process becomes important when ram
pressure, which depends on the satellite velocity and gas density of
the host halo, overcomes the restoring gravitational force generated
by the satellite's mass distribution \citep{McCarthy2008}. In MW-mass
haloes ram pressure does not overcome the gravitational restoring
force of LMC-mass dwarfs \citep{Simpson2017}, but ram pressure
increases rapidly with host halo mass, since the satellites of more
massive hosts are moving more rapidly inside a denser gas medium. 

\vspace{-.3cm}
\section{LMC-mass dwarfs in MW-mass hosts}
\label{sect:MW}
\begin{figure}
    \plotone{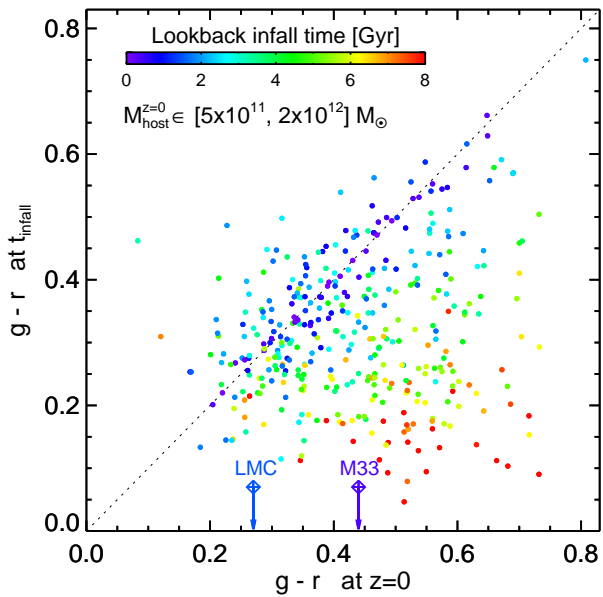}
	\caption{ The colour evolution of LMC-mass dwarfs that are
          satellites of MW-mass haloes. We show the $g-r$ colour at
          infall versus the $g-r$ colour at $z=0$. Each symbol
          corresponds to an LMC-mass satellite, with the 
          colour indicating the lookback time to infall (see legend). The
          two vertical arrows show the present-day colours of the LMC
          and M33 and the arrows are coloured according to the
          estimated infall time of the satellites.
 	}
    \label{fig:Color_infall}
    \vspace{-.3cm}
\end{figure}
\begin{figure}
 	\plotone{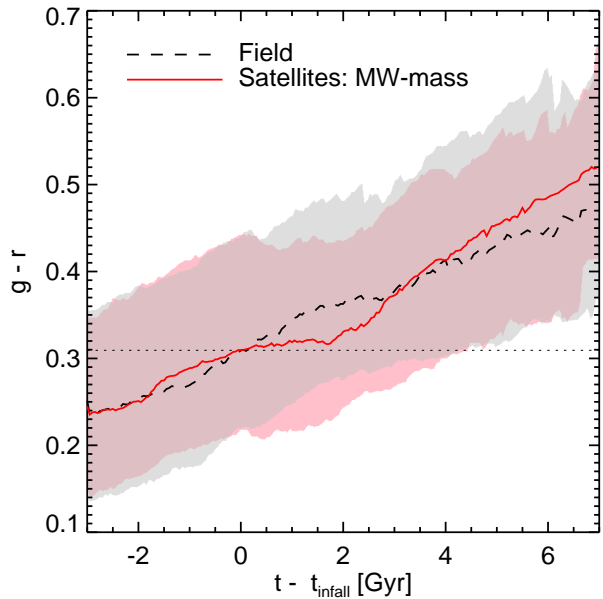}
 	\caption{ The evolution of the colour distribution for two
          samples of LMC-mass dwarfs: satellites of MW-mass hosts (in
          red) and a control sample of field galaxies (in black). The
          latter was obtained by pairing each satellite at infall with
          a field dwarf of the same colour. The evolution is expressed
          as a function of time after infall for each satellite
          galaxy, with infall time defined as $t=0$, and $t<0$ and
          $t>0$ corresponding to before and after infall 
          respectively. The dashed and solid lines correspond to the
          medians of the two distributions, while the shaded regions
          show the 16th and 84th percentiles. The dotted horizontal
          line is shown for reference, and corresponds to the median
          colour value at $t=0$. 
 	}
 	\label{fig:Color_evo}
    \vspace{-0.3cm}
\end{figure}
In this section we investigate in more detail the properties and evolution
of LMC-mass satellites around MW-mass host haloes. In particular, we
investigate differences between infall time, pericenter and evolution
of red and blue LMC-mass dwarfs, and relate these properties to the
brightest Galactic satellite, the LMC.

\vspace{-.2cm}
\subsection{Abundance}
We first study the abundance of LMC-mass dwarfs around MW-mass
haloes. As discussed in \refsec{sect:simul}, we found 381 \eagle{}
LMC-mass satellite galaxies residing in MW-mass hosts with masses
in the range, $M_{\rm 200} \in [0.5, 2] \times 10^{12}\Msun$. We
split this sample into three subsets according to the host
halo mass, and for each subset we calculate the fraction of hosts that
have at least one LMC-mass satellite as a function of the satellite's
stellar mass. The outcome is shown in \reffig{fig:Pdf_mw}. The
probability of finding a massive dwarf depends primarily on the host
halo mass, and, for a fixed host halo mass, it decreases with
increasing stellar mass of the satellite
\citep{Boylan-Kolchin2011,Busha2011,Cautun2014a}. 

Satellite dwarfs with a stellar mass of $2.7\times 10^{9}\Msun$, which
corresponds to the LMC, are very rare (4\%) in haloes with
$M_{\rm 200} \in [0.5, 0.8]\times10^{12}\Msun$ and somewhat more
common ($16\%$) in haloes with
$M_{\rm 200} \in [0.8, 1.5]\times10^{12}\Msun$, in agreement with
previous theoretical and observational studies
\citep[e.g.][]{Boylan-Kolchin2011,Robotham2012,Guo2013}. The presence
of such a massive satellite around the MW imposes a lower limit on the
MW halo mass, such that masses as low as
$M_{\rm 200} \sim 0.5\times10^{12}\Msun$ are unlikely. The most
stringent constraint comes when the SMC is also included, which is
itself unexpectedly massive, to suggest a MW halo mass larger than
$1.0\times 10^{12}\Msun$ with 90\% confidence \citep{Cautun2014a}.

\vspace{-.2cm}
\subsection{Colour evolution}
\begin{figure*}
    \vspace{-0.5cm}
	\plottwo{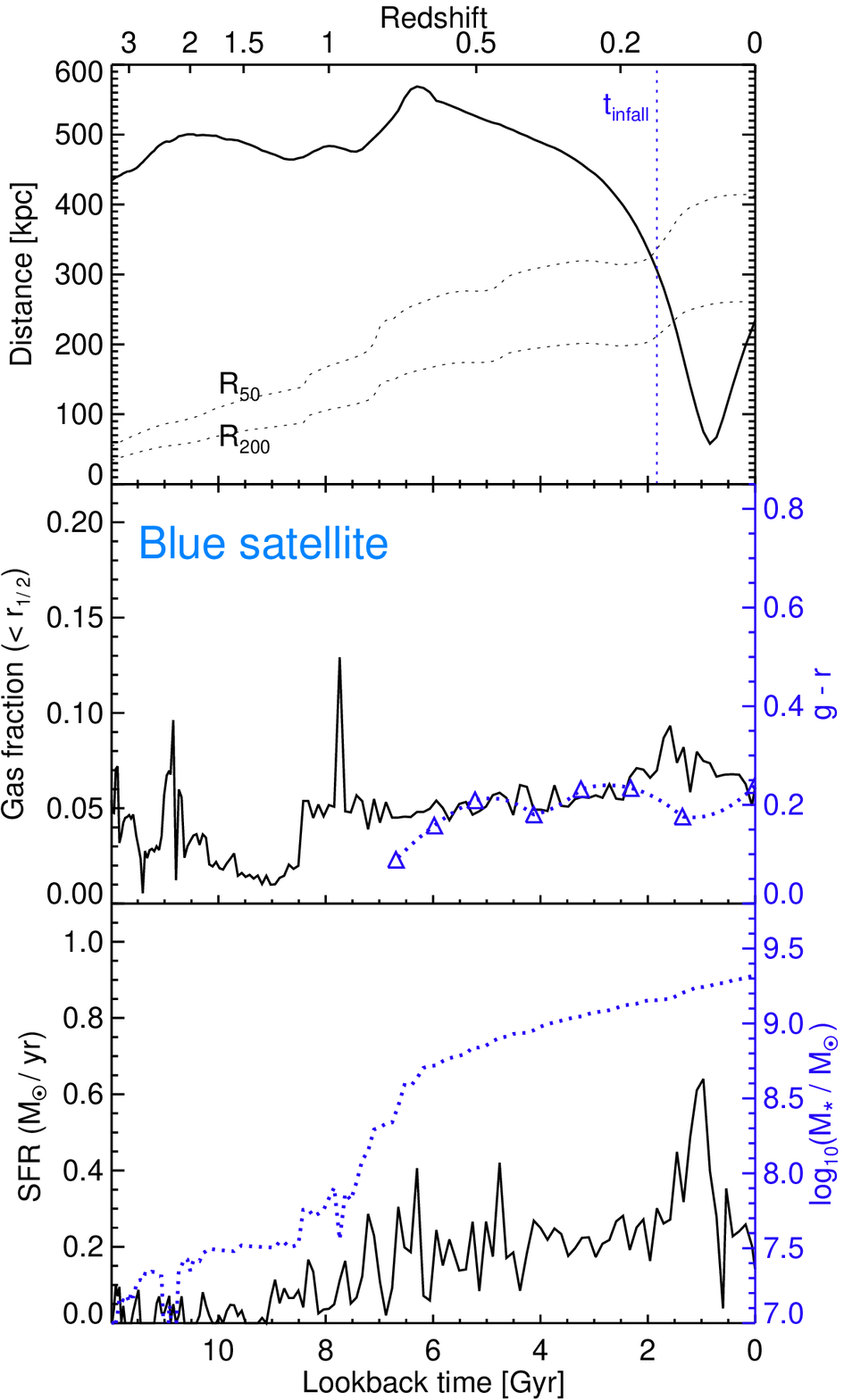}{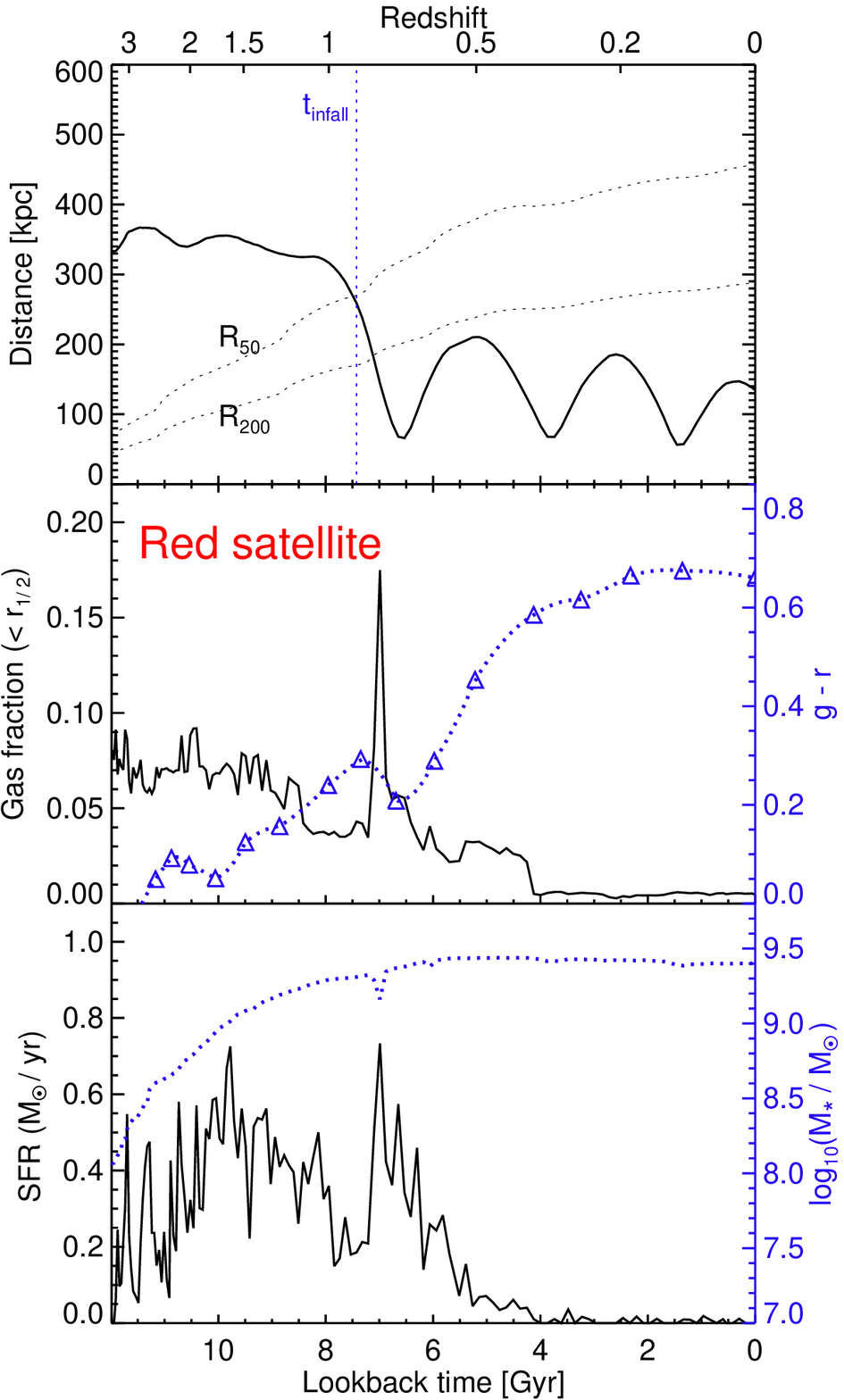}
	\caption{The formation history of two LMC-mass dwarfs which by
          the present have become satellites of two MW-mass haloes. The
          left and right columns show LMC-mass dwarfs that, at
          $z=0$, have blue and red $g-r$ colours, respectively. \textit{Top
            row:} the distance between the dwarf and its present day
          host halo, with dotted lines indicating the evolution of
          the host radii, $R_{200}$ and $R_{50}$, and the vertical
          dashed line indicates the moment of infall into the MW-mass
          host. \textit{Middle row:} the evolution of the LMC-mass
          dwarf's gas fraction (left axis) and $g-r$ colour (right
          axis). 
    Colours are available only for galaxies with
    $M_\star>5\times10^8\Msun$ and for a limited number of
    redshifts. \textit{Bottom row:} the evolution of the LMC-mass
    dwarf SFR (left axis) and stellar mass (right axis). 
	}
    \label{fig:examples}
    \vspace{-0.3cm}
\end{figure*}
We show in \reffig{fig:examples} the evolution of two LMC-mass
galaxies which by $z=0$ have become satellites of MW-mass haloes. The one
shown on the left is analogous to our LMC: it has a very blue
$g-r$ colour, is actively forming stars and it recently passed its
first pericentre, having fallen into its host MW halo only $2$ Gyrs
ago. We contrast this blue dwarf with a red LMC-mass
satellite, which is shown in the right panels of
\reffig{fig:examples}. The two examples offer the opportunity
to highlight both similarities and differences between blue and red
satellites. The discussion which follows is based on investigating a
larger sample of LMC-mass satellites and summarizes the typical
behaviour seen for the majority of objects (some of the properties are
studied in more detail in later figures). 

On average, red satellites have fallen in a longer time ago and, in
many cases, had a smaller gas fraction at infall than blue satellites
mostly due to self-quenching. The latter is not the case for the red
dwarf shown in \reffig{fig:examples}, which at infall had a similar
gas fraction as the example blue dwarf. Once accreted, many satellites
experience an episode of gas compression, which leads to increased
star formation \citep{Dressler1983,Zabludoff1996,Sabatini2005}.
This phenomenon has also been seen in cluster galaxies in
the \textsc{illustris} simulation \citep{Mistani2016}, which has a different
treatment of baryonic physics from \eagle{}. This episode
typically occurs in gas rich dwarfs shortly after entering the
halo and, in the examples shown in \reffig{fig:examples}, it takes place
at a lookback time of $1.5$ and $7.0$ Gyrs for the blue and red
satellites, respectively. The typical gas compression is similar to
that seen in the blue dwarf example, but there is also a significant
fraction of the population that undergoes very strong gas compression
similar to the one seen in the red dwarf example. This effect,
enhanced star formation due to gas compression, is the reason why 
LMC-mass satellites in MW-mass hosts have, on average, both higher
sSFR and blue colours than the field population (see Figs
\ref{fig:Pdf_sfr} and \ref{fig:Pdf_color}).

The two examples in \reffig{fig:examples} highlight another process that
affects the evolution of dwarf galaxies: mergers with other dwarfs \citep{Deason2014}.
Both dwarfs had at least one merger with another dwarf galaxy, which took
place at a lookback time of $7$ and $9$ Gyrs for the blue and red LMC-mass
analogues, respectively. The merger can be inferred from the small wiggles
in the distance plot shown in the top panel of \reffig{fig:examples}, which
are due to the relative orbital motion of the merging dwarfs. For the blue
satellite, the merging dwarf is massive and its disruption leads to a sudden
increase in the stellar mass of the LMC-mass progenitor (see dashed line in
the bottom-left hand panel of \reffig{fig:examples}). The merger leads to
rapid gas compression and enhanced star formation. In contrast, the progenitor
of the red LMC-mass dwarf experiences a lower mass merger and its imprint on
both the gas fraction and SFR is less pronounced, with possibly enhanced SFR
around 10 Gyrs ago.

\begin{figure}
    \vspace{-0.2cm}
	\plotone{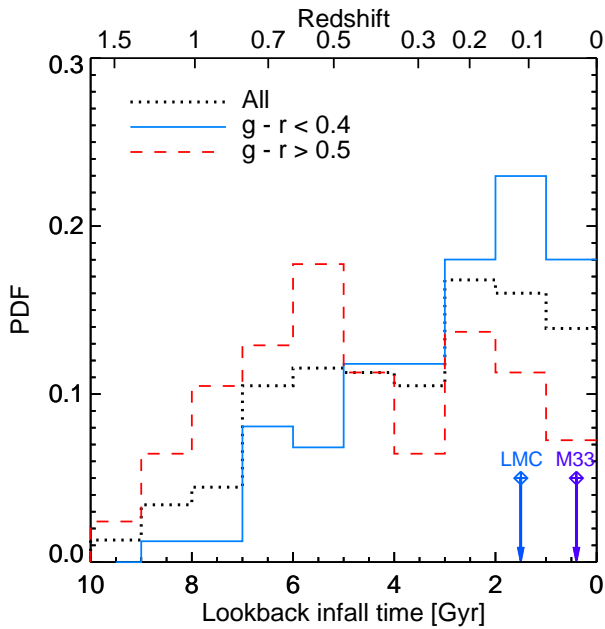}
	\caption{ The distribution of infall times for the LMC-mass
          dwarfs that are satellites of MW-mass haloes. Shown are the
          distributions for the entire sample (dotted line) and for the
          sample split according to $z=0$ colour: $g-r<0.4$ (solid
          line) and $g-r>0.5$ (dashed line). The two colour-selected
          subsamples correspond to roughly a third of the full
          sample. The two vertical arrows show the estimated infall
          times of the LMC and M33 (the M33 estimate is highly
          uncertain; see discussion in main text). 
	}
	\label{fig:Pdf_infall}
    \vspace{-0.5cm}
\end{figure}

In order to understand the colour evolution of LMC-mass
satellites better, we show the correlation between $g-r$ colour at infall and
at the present time in \reffig{fig:Color_infall}. Each point corresponds
to an LMC-mass satellite in a MW-mass host, with colour reflecting 
the lookback time to infall. At infall, most dwarfs are blue,
i.e. $g-r<0.6$, with only 4 out of the 381 dwarfs that are red. Many of
the galaxies that fell in recently ($<2$ Gyrs) have, on average, at
$z=0$, slightly bluer colours than at infall, which is due to the
enhanced star formation that takes place in these dwarfs when they first
enter a MW-mass host halo. In contrast, galaxies accreted between 3 to
6 Gyrs ago are significantly redder than at infall and have typically
experienced a colour change $\Delta(g-r)\simeq 0.15$. Galaxies
accreted more than 6 Gyrs ago show the largest reddening,
corresponding to a colour change since infall of $\Delta(g-r)\simeq
0.3$ or higher. Interestingly, while the satellites accreted the
earliest show the largest change in colour, they were, on average, 
very blue at infall and thus are not necessarily classified as red,
i.e. as having $g-r > 0.6$. The red satellites have a large
distribution of accretion ages, being a mixture of dwarfs accreted
long ago, and recently accreted objects whose colour at infall
was slightly bluer than $g-r=0.6$. 

The LMC is estimated to have been accreted into the MW about 1.5 Gyrs
ago \citep{Patel2017} and its very blue colour is
consistent with this prediction. In contrast, the orbit of M33 is much
more uncertain, with predicted infall times in the literature varying
from 0.4 Gyrs \citep{Patel2017} to more than 4 Gyrs
\citep{Putman2009,McConnachie2009}. The former are based on proper
motions for both M33 and M31, but an early accretion scenario only
includes a small region of the allowed proper motion space. The latter
use the warped HI disc \citep{Putman2009} and the faint stellar
structure surrounding M33 \citep{McConnachie2009} as evidence of a
past close encounter between M33 and M31, suggesting that M33 was
accreted at least several Gyrs ago. The $g-r$ colour of M33 is
unlikely to distinguish between the two scenarios (see
\reffig{fig:Color_infall}), since its present day colour is consistent
with both late and early accretion, with the latter option being
acceptable if M33 was very blue when it fell into M31. Curiously, the
M33 star formation history has a prominent peak around 2 Gyrs ago
\citep{Williams2009} that could correspond to enhanced star formation
due to gas compression within ${\sim}1$ Gyr after infall into M31 (see
discussion of \reffig{fig:examples}). This hypothesis would favour the
early accretion scenario.

\reffig{fig:Color_evo} contrasts the colour evolution of LMC-mass
satellites around MW-mass haloes with that of similar dwarfs in the
field. The latter were selected by assigning to each satellite at
infall a field counterpart of the same colour. The figure shows the
evolution from 3 Gyrs before infall to 7 Gyrs after infall. Before
infall, we find a close match in the evolution of the satellite and
field samples. Since these two samples were matched to have the same
colours at infall, this indicates that MW-mass haloes do not affect
the evolution of LMC-mass dwarfs outside $R_{50}$. After infall, for
the next ${\sim}3$ Gyrs, the satellites are bluer on average than they
would have been had they stayed in the field. As we discussed above,
this is due to the enhanced star formation triggered by gas
compression. Interestingly, the colour of the satellites remains the
same for up to 2 Gyrs after infall, after which it starts to redden
faster than in their field counterparts. By 6 Gyrs, the satellites are
$\Delta(g-r)\simeq 0.2$ redder than at infall, and
$\Delta(g-r)\simeq 0.05$ redder than if they would have remained in
the field.

\begin{figure}
    \vspace{0.3cm}
	\plotone{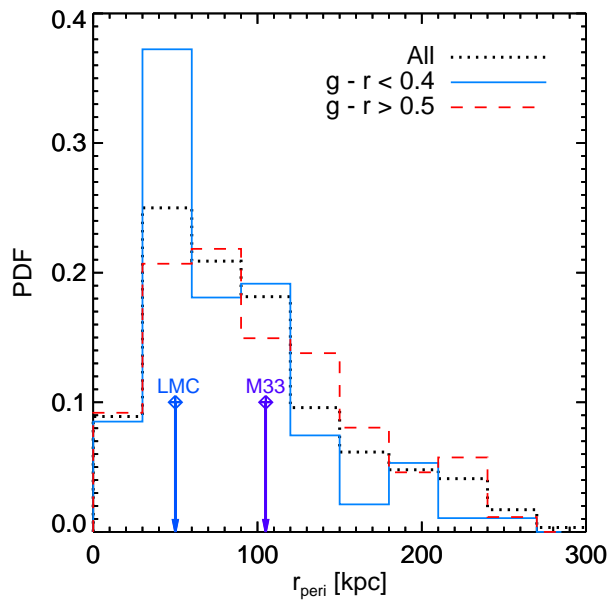}
	\caption{ The distribution of first pericentric distances for
          LMC-mass dwarfs that are satellites of MW-mass haloes. As in
          \reffig{fig:Pdf_infall}, we show the full sample (dotted
          line) and $g-r$ colour selected subsamples: $g-r<0.4$ (solid
          line) and $g-r>0.5$ (dashed line). The two vertical arrows
          show the estimated first pericentre distance of the LMC and
          M33.  }
	\label{fig:Pdf_peri}
    \vspace{-0.4cm}
\end{figure}

The two examples in \reffig{fig:examples}, as well as
\reffig{fig:Color_infall}, showcase the importance of infall time:
early accreted dwarfs are redder than late accreted ones. To
study the dependence between infall time and the present
day $g-r$ colour statistically, we split the LMC-mass satellites of MW-mass haloes
into two subsets according to their $z=0$ colour: the reddest third,
corresponding to $g-r > 0.5$, and the bluest third, where $g-r <
0.4$. As the name suggests, each subset contains roughly one third of
the full sample. \reffig{fig:Pdf_infall} shows the distribution of
lookback times to infall for these two subsets, with $t=0$
corresponding to the present day. We find a strong trend of the
present day colour with infall time, with the reddest third subset
having earlier infall times on average. In contrast, the bluest third
subset was generally accreted more recently, with most objects having
fallen into their MW hosts less than 7 Gyrs ago. 

It is intriguing to compare the infall time of LMC-mass satellites
with that of lower mass satellites of MW-mass haloes. \citet{Shao2017}
studied the distribution of infall times for the brightest 11
satellites of \eagle{} MW-mass haloes to find that most such dwarfs
were accreted between 8 and 10 Gyrs ago, with only $40\%$ of objects
having lookback times to infall below 7 Gyrs. In contrast, $50\%$ of
the LMC-mass satellites were accreted less than 3.5 Gyrs ago. Due to
their higher total mass, LMC-mass dwarfs experience strong dynamical
friction and thus sink towards the halo centre, where they end up
being tidally disrupted and possibly merging with the central
galaxy. 

In \reffig{fig:Pdf_peri} we investigate if the colour evolution of
LMC-mass satellites depends on their orbit. We plot the distribution
of first pericentre distances for all the sample, as well as for the
bluest-third and reddest-third subsets. The distribution peaks around
${\sim}50\kpc$ and drops sharply for smaller distances, while for
distances larger than $50\kpc$ there is a more gradual decrease. We
find a clear difference between the pericentre distances of the
bluest-third and those of the reddest-third subset, with the former
typically closer to the halo centre. This suggests that, on
average, satellites that get close to the central galaxy are more
likely to experience gas compression and thus form more stars.

\reffig{fig:Pdf_peri} also shows the predicted LMC and M33 pericentres
of 50 and $105\kpc$, respectively \citep{Patel2017}. These
measurements are in good agreement with the simulation predictions. In
particular, the LMC pericentre is near the peak of the
distribution. Interestingly, the \citet{McConnachie2009} scenario of a
close encounter between M33 and M31 in order to explain the warped HI
disc and the extended stellar structure of M33, requires a pericentric
passage of ${\sim}50\kpc$; this value is favoured more by the \eagle{}
data than the ${\sim}100\kpc$ pericentre suggested by the M31 and M33
proper motions.

\vspace{-0.4cm}
\section{Conclusions}
\label{sect:conclusions}
We have investigated the properties of LMC-mass dwarf galaxies in the
main \eagle{} cosmological hydrodynamics simulations of galaxy
formation. By LMC-mass dwarfs we mean the population of galaxies in
the simulation that have a stellar mass similar to the LMC, i.e. in
the range $[1,4]\times10^9\Msun$. \eagle{} is well suited to this
study because of its rare combination of high resolution and large
volume, and because it produces a population of galaxies with
realistic masses, sizes, star formation rates, colours and gas
content. To understand the effects of environment, the LMC-mass dwarfs
were split into satellite and field galaxy samples. The former are
dwarfs which are inside the halo of a brighter galaxy while the latter
are central galaxies. In order to focus on objects similar to the LMC and
M33, which are the brightest satellites of the MW and M31,
respectively, we selected a further subset of LMC-mass satellites
hosted by MW-mass haloes.

\noindent ~~~~ Our main conclusions are summarised as follows:
\\[-.55cm]
\begin{enumerate}
 	\item Field LMC-mass dwarfs reside in haloes with
          $M_{200}\sim2\times10^{11}\Msun$ and have a stellar-to-halo
          mass ratio of $1.03^{+0.50}_{-0.31}\times10^{-2}$, in
          agreement with abundance matching estimates (see Figs
          \ref{fig:Star_m200} and \ref{fig:Pdf_mf}). Furthermore,
          LMC-mass centrals that have a SMC-mass satellite reside in
          haloes 1.3 times more massive than the typical LMC-mass
          dwarf. This suggests that the LMC halo mass at infall was
          relatively high; \eagle{} predicts $M_{200} =
          3.4^{+1.8}_{-1.2} \times 10^{11} \Msun$ (68\% confidence
          interval). 
 	\\[-.3cm] 
 	\item In agreement with observations, the $g-r$ colour distribution
          is bimodal with the red mode consisting mainly of LMC-mass
          satellites of massive groups and clusters (see
          \reffig{fig:Pdf_color}). Field galaxies have a unimodal
          colour distribution and are mostly blue; only $15\%$ of
          them are red, i.e. they have $g-r>0.6$. The quenching of
          field dwarfs is predominantly driven by self-quenching. 
 	\\[-.3cm]
 	\item The fraction of satellites that are red increases
          rapidly with host mass, from $10\%$ for MW-mass hosts, to
          $50\%$ for hosts with $M_{200}=5\times10^{12}$, and then to
          over $90\%$ for hosts with $M_{200}>3\times10^{13}$ (see
          \reffig{fig:Pdf_red}). 
 	\\[-.3cm]
      \item The quenching timescale, defined as the time after infall
        when half of the satellites have acquired red colours, varies
        strongly with host halo mass, with values of ${>}5$, $5$ and
        $2.5$ Gyrs for hosts with masses, $M_{200}\sim10^{12}$,
        $10^{13}$ and $10^{14}\Msun$, respectively
        (\reffig{fig:Pdf_red_evo}). It indicates that the dominant
        quenching process varies with host halo mass, from starvation
        in the case of MW-mass hosts to ram pressure stripping for
        clusters.
 	\\[-.3cm]
 	\item LMC-mass satellites hosted by MW-mass haloes show
          enhanced star formation and bluer $g-r$ colours than both
          the field and the overall satellite population (see Figs
          \ref{fig:Pdf_sfr} and \ref{fig:Pdf_color}). Shortly after
          accretion into the MW-mass host, the dwarfs experience gas
          compression that leads to an episode of increased star
          formation. 
 	\\[-.3cm]
 	\item The prevalence of LMC-mass satellites in MW-mass haloes
          depends primarily on halo mass. The presence of the LMC
          in MW and M33 in M31 suggests that the two giant galaxies
          reside in haloes more massive than ${\sim}10^{12}\Msun$ (see
          \reffig{fig:Pdf_mw}).
 	\\[-.3cm]
 	\item After infall into MW-mass haloes, LMC-mass dwarfs
          have slightly bluer colours for ${\sim}2$ Gyrs, after which
          they quickly redden, with on average $\Delta(g-r)=0.2$ and
          $0.4$ after 6 and 8 Gyrs from infall, respectively (see
          \reffig{fig:Color_infall}).
  	\\[-.3cm]
  	\item More than half of the LMC-mass satellites of MW-mass hosts were accreted less than 3.5 Gyrs and most (${\sim}70\%$) within the last 5 Gyrs (see \reffig{fig:Pdf_infall}). In contrast, less than 30\% of satellites with similar mass to the classical MW dwarfs were accreted within the last 5 Gyrs \citep{Shao2017}.
\end{enumerate}

One of the goals of this paper has been to understand better the
processes that dominate the formation of LMC-mass dwarfs, with
particular emphasis on the LMC and M33 galaxies. 
The orbit of M33 is not very well constrained because of large
uncertainties in the proper motions of M33 and M31. Currently,
the most likely scenario inferred from proper motion data is that M33
fell into the M31 halo only recently, ${\sim}0.4$ Gyrs ago, and is on
an elongated orbit with a first pericentre distance of $100\kpc$
\citep{Patel2017}. However, this seems inconsistent with the warped HI
disc and the extended stellar distribution around M33, which could
naturally be explained by a close encounter with M31, e.g. with a
pericentre distance of ${\sim}50\kpc$ about $3$ Gyrs ago
\citep{McConnachie2009}. Our results favour the second possibility
because: 1) the enhanced star formation rate in M33 around
2~Gyrs ago, which we found to arise naturally from gas compression
after infall into the larger halo; and 2) the higher likelihood of a
pericentric distance of $50\kpc$, which is twice as likely as the
larger values expected in the very recent infall scenario.

The LMC, whose orbit is better constrained than the M33 one, is thought to
have been accreted around 1.5 Gyrs ago \citep{Patel2017} and both its
current enhanced star formation and very blue colours can be explained
by gas compression upon entry, which we found to be common among the
recently accreted satellites of MW-mass haloes. The distribution of
infall times suggest that LMC-mass satellites of MW-mass haloes have
short lifetimes, with dynamical friction rapidly causing their orbit
to decay towards their host centre. 

\vspace{-.5cm}
\section*{Acknowledgements}
We thank the anonymous referee for detailed comments that have helped us improve the paper.
SS, MC and
CSF were supported by the Science and Technology Facilities Council
(STFC) [grant number ST/F001166/1, ST/I00162X/1, ST/P000541/1]. AD is
supported by a Royal Society University Research Fellowship. This work
used the DiRAC Data Centric system at Durham University, operated by
ICC on behalf of the STFC DiRAC HPC Facility (www.dirac.ac.uk). This
equipment was funded by BIS National E-infrastructure capital grant
ST/K00042X/1, STFC capital grant ST/H008519/1, and STFC DiRAC
Operations grant ST/K003267/1 and Durham University. DiRAC is part of
the National E-Infrastructure. 

\vspace{-.7cm}
\bibliographystyle{mnras}
\bibliography{bibliography}
\label{lastpage}
\end{document}